%% file: HBTL.tex
\documentclass[envcountsame,envcountreset,envcountsect]{llncs}
\usepackage{graphicx}

\usepackage{volker}
\usepackage{amsmath} 
\usepackage{amssymb} 
\usepackage{pst-node}

\usepackage{enumerate,multicol,theorem}

\newcommand{\ignore}[1]{}
\newcommand{\toappendix}[1]{\marginpar{!}{\scriptsize }}

\spnewtheorem{mydefinition}[theorem]{Definition}{\bfseries}{}

\begin{document}

\title{Hybrid Branching-Time Logics\thanks{An extended abstract of this paper was presented at the International Workshop on Hybrid Logics (HyLo 2007).}} 
\author{Volker Weber}
\institute{Fachbereich Informatik, Universit\"at Dortmund\\44221 Dortmund, Germany\\ \email{\texttt{volker.weber@udo.edu}}}
\maketitle

\begin{abstract}
Hybrid branching-time logics are introduced as extensions of \ctl-like logics with state variables and the downarrow-binder. Following recent work in the linear framework, only logics with a single variable are considered. The expressive power and the complexity of satisfiability of the resulting logics is investigated.

As main result, the satisfiability problem for the hybrid versions of several branching-time logics is proved to be \TWOEXPTIME-complete. These branching-time logics range from strict fragments of  \ctl to extensions of \ctl that can talk about the past and express fairness-properties. The complexity gap relative to \ctl is explained by a corresponding succinctness result.

To prove the upper bound, the automata-theoretic approach to branching-time logics is extended to hybrid logics, showing that non-emptiness of alternating one-pebble B\"uchi tree automata is \TWOEXPTIME-complete.
\end{abstract}

\input{introduction}

\section{Preliminaries}\label{sec:defs}

The basic definitions of branching-time logics and B\"uchi tree automata are presented in this section. As both formalism are defined with respect to infinite trees, we start by defining these structures.

Let $D$=\{1,\ldots,k\} be a finite set of directions for some $k\in\mathbb{N}$.
An \emph{infinite $D$-tree} is a prefix-closed set $T\subseteq D^*$, i.e., whenever $x\cdot c\in T$ where $x\in D^*$ and $c\in D$, then also $x\in T$. 
The empty string $\varepsilon$ is the \emph{root} of $T$ and  for all $c\in D$, $x\cdot c$ is called a \emph{child} of the node $x$.
A \emph{path} $\pi$ in $T$ is a prefix-closed minimal set $\pi\subseteq T$, such that for every $x\in\pi$, there is a unique $c\in D$ with $x\cdot c\in\pi$.
We use ``$\leq$'' to denote the descendant-relation on $T$, i.e., $x<y$ if and only if $y$ is a strict descendant of $x$. Note that this order is partial as nodes in different branches are incomparable.
The branching degree $d(x)$ is the number of children of a node $x$. We only consider \emph{$k$-ary trees}, where $d(x)=k$ for every node $x$, and refer to them as trees in the following. 

A \emph{labeled tree} over a finite alphabet $\Sigma$ is a pair $(T,V)$ where $T$ is a tree and $V:T\rightarrow\Sigma$ assigns a symbol from $\Sigma$ to every node of $T$. We are mainly interested in the case in which $\Sigma=2^{\prop}$ for some set $\prop$ of propositions. Such trees, usually referred to as \emph{computation trees}, result for example from the unwinding of Kripke structures (see, e.g., \cite{KupfermanV06}). In the following, we identify $(T,V)$ with $T$.

\input{logic_def}
\input{automata_def}

\input{expressiveness}

\input{satisfiability}

\input{automata}
\input{conclusion}

\bibliographystyle{abbrv}
\bibliography{HBTL}

\end{document}

%% file: introduction.tex
\section{Introduction}\label{sec:introduction}

Hybrid logics are extensions of modal logic that allow to refer to individual states of a model. They aim at extending the expressive power of modal logics, without losing their nice properties such as decidability. Hybrid logics have been researched quite intensively during the last years. Their applications range from verification to reasoning about semistructured data \cite{FranceschetR05}. See \cite{ArecesC07} for a recent survey and an introduction to hybrid logic.

On the technical side, the aims of hybrid logic can be achieved by adding nominals and state variables, corresponding to the first-order concepts of constants and variables. Nominals are an additional kind of atomic symbols which are true in exactly one state in a model, and therefore name this state. Nominals are fixed with the model, whereas the assignment of states to state variables can be changed by quantification. To preserve the local perspective of modal logic, the quantifier considered the most is the downarrow-operator, first introduced in \cite{Goranko94} and denoted $\mydownarrow$, which binds a state variable to the current state.

Satisfiability of hybrid \mydownarrow-languages is undecidable with respect to arbitrary Kripke-structures \cite{ArecesBM99} and only non-elementarily decidable if the class of models is restricted to trees \cite{mundhenkSSW05} or linear structures \cite{FranceschetRS03}, i.e., to those models important in verification. These results initiated research on decidable fragments and fragments of lower complexity \cite{CateF05,SchwentickW07}.

In \cite{SchwentickW07}, Schwentick and W. considered bounded-variable fragments of hybrid \mydownarrow-languages in the linear framework. While complexity of the two-variable fragment is already as bad as for the unbounded language, satisfiability of the one-variable fragment is \EXPSPACE-complete. Furthermore, the one-variable fragment has the full expressive power of first-order logic.

The aim of this paper is to extend this successful approach to the branching-time framework. While the logic of \cite{SchwentickW07} could also be interpreted over trees, we believe that hybrid extensions of classical branching-time logics like \ctl are a more convenient formalism to reason about trees.

Our main result is that satisfiability for the one-variable-fragment of hybrid $\ectl+$Past is \TWOEXPTIME-complete. The lower bound is already achieved for the logic containing only the next \myX and future \myF modalities and is explained by a corresponding succinctness result. The upper bound is by a reduction to non-emptiness of alternating one-pebble B\"uchi tree automata, a problem that we prove to be \TWOEXPTIME-complete as well. Furthermore, we study the expressive power of hybrid branching-time logics and show, for example, that the one-variable-fragment of hybrid \ctl is strictly more expressive than \ctl.

Section~\ref{sec:defs} gives the basic notions of branching-time logics and introduces their hybrid extension. It also contains the definition of alternating one-pebble B\"uchi tree automata. Section~\ref{sec:expressiveness} is concerned with the expressive power of hybrid branching-time logics, which are compared with classical branching-time logics and logics with the \myN-operator (``from now on'') of \cite{LaroussinieS95}. The complexity and succinctness results can be found in Section~\ref{sec:satisfiability}, those on tree automata in Section~\ref{sec:automata}. We give some directions for further research in Section~\ref{sec:conclusion}.

%% file: logic_def.tex
\subsection{Branching-Time Logics}

We briefly recall the basic notions of branching-time logic, starting from the logic $\ectl+$Past which extends \ctl in two ways: First, by adding the $\Finf$-modality for ``infinitely often in the future'' to express fairness properties. Secondly, by introducing past modalities $\myY$ for ``previous'' and $\myS$ for ``since''.

$\ectl+$Past-formulas are defined by the following grammar:
$$\varphi::=p\mid\neg\varphi\mid\varphi\wedge\varphi\mid\ex\varphi\mid\myE\varphi\myU\varphi\mid\myA\varphi\myU\varphi\mid\efinf\mid\myY\varphi\mid\varphi\myS\varphi$$
where $p\in\prop$. We use the usual abbreviations $\true$, $\false$, $\varphi\vee\varphi$, $\varphi\rightarrow\varphi$, and
\begin{center}
	\begin{tabular}{rclrclrcl}
	 $\ef\varphi$ & $:=$ & $\myE\true\myU\varphi\qquad\qquad$ & $\eg\varphi$ & $:=$ & $\neg\af\neg\varphi\qquad\qquad$ & $\aginf\varphi$ & $:=$ &  $\neg\efinf\neg\varphi$\\
	 $\af\varphi$ & $:=$ & $\myA\true\myU\varphi\qquad\qquad$ & $\ag\varphi$ & $:=$ & $\neg\ef\neg\varphi\qquad\qquad$ &  $\afinf\varphi$ & $:=$ &  $\ag\af\varphi$\\
	 $\ax\varphi$ & $:=$ & $\neg\ex\neg\varphi$ & $\myP\varphi$ & $:=$ & $\true\myS\varphi$ & $\eginf\varphi$ & $:=$ &  $\ef\eg\varphi$
	\end{tabular}
\end{center}

The semantics of $\ectl+$Past-formulas is defined with respect to a computation tree $T$ and a node $n$ of $T$:
\begin{itemize}
	\item $T,n\models p$ iff $p\in V(n)$
	\item $T,n\models \neg\varphi$ iff $T,n\not\models\varphi$
	\item $T,n\models \varphi\wedge\psi$ iff $T,n\models\varphi$ and $T,n\models\psi$
	\item $T,n\models \ex\varphi$ iff there exists a $c\in D$, such that $T,n\cdot c\models\varphi$
	\item $T,n\models \myE\varphi\myU\psi$ iff there exists a path $\pi$ with $n\in\pi$, such that there is a descendant $n'\in\pi$ of $n$ with $T,n'\models\psi$ and for all nodes $x\in\pi$ with $n\leq x< n'$, we have $T,x\models\varphi$ 
	\item $T,n\models \efinf\varphi$ iff there exists a path $\pi$ with $n\in\pi$, such that there are infinitely many descendants $n'\in\pi$ of $n$ with $T,n'\models\varphi$ 
	\item $T,n\models \myA\varphi\myU\psi$ iff for all paths $\pi$ with $n\in\pi$, there is a descendant $n'\in\pi$ of $n$ with $T,n'\models\psi$ and for all nodes $x\in\pi$ with $n\leq x< n'$, we have $T,x\models\varphi$ 
	\item $T,n\models\myY\varphi$ iff $T,n'\models\varphi$ with $n=n'\cdot c$ for some $c\in D$
	\item $T,n\models\varphi\myS\psi$ iff there exists an ancestor $n'$ of $n$, such that $T,n'\models\psi$ and for all nodes $x$ with $n'< x\leq n$, we have $T,x\models\varphi$
\end{itemize}

Two formulas $\varphi$ and $\psi$ are \emph{equivalent}, if $T,\varepsilon\models\varphi\iff T,\varepsilon\models\psi$ for all computation trees $T$, i.e., we only consider initial equivalence as we want to compare the expressive power of logics with and without past modalities \cite{LaroussinieS95}.

We consider several fragments of $\ectl+$Past and denote them by \bt{C}, where $C$ is the set of temporal operators allowed. To give some examples:
\begin{itemize}
	\item \bxf is the logic \ub of \cite{Ben-AriPM83},
	\item \bxu is the well known logic \ctl \cite{ClarkeE81},
	\item \bxufinf was introduced in \cite{EmersonH86} as \ectl, and
	\item \bxuys is \pctl, i.e., \ctl+Past.
\end{itemize}

In all these logics, future temporal operators occur only immediately in the scope of the path quantifiers $\myE$ and $\myA$. Opposed to this, the branching time logic \ctlstar from \cite{EmersonH86} allows Boolean combinations and nesting of these operators. 

\subsection{Hybrid Branching-Time Logics}

We extend branching-time logics with hybrid machinery along the lines of \cite{SchwentickW07}. I.e., we use only one state variable $x$ and consider only a single nominal $root$.
 
\begin{mydefinition}
Given a set $C\subseteq\{\myX,\myF,\myU,\Finf,\myY,\myP,\myS\}$ of modalities and a branching-time logic \bt{C}, the formulas of the corresponding \emph{hybrid branching-time logic} \hb{C} are those of \bt{C} and additional
$$\mydownarrowx\varphi\mid x\mid \myatx\varphi\mid root\mid \atroot\varphi,$$
where $\varphi$ is a \hb{C}-formula and $x$ is the only state variable.

The semantics of hybrid branching-time formulas is defined with respect to a computation tree $T$ and two nodes $n,m$ of $T$, where $n$ is the current node and $m$ is the node assigned to the state variable $x$: 
\begin{center}
	\begin{tabular}{lcllcl}
		$T,n,m\models \mydownarrowx\varphi$ & iff & $T,n,n\models \varphi\qquad\qquad$ & $T,n,m\models x$ & iff & $n=m$\\
		$T,n,m\models \myatx\varphi$ & iff & $T,m,m\models \varphi\qquad\qquad$ & $T,n,m\models root$ & iff & $n=\varepsilon$\\
		$T,n,m\models \atroot\varphi$ & iff & $T,\varepsilon,m\models \varphi\qquad\qquad$
	\end{tabular}
\end{center}
and the semantics of classical branching-time logic are extended in the obvious way, i.e., the state variable is not affected.

A formula $\varphi$ is called \emph{satisfiable} if there is a computation tree $T$ and nodes $n,m$ such that $T,n,m\models\varphi$. 
\end{mydefinition}
\begin{remark}[Using hybrid machinery] We give two examples on how hybrid branching-time logics work. The reader will find both patterns again in the proofs given in this paper.

Hybrid branching-time logics can reason about the past without using past modalities. The past formula $\myP\varphi$ can be expressed as $\mydownarrowx\atroot\ef(\ef x\wedge\varphi)$, for example. This illustrates how a finite prefix of a path can be fixed.

Moreover, they can easily compare two nodes. The property that there are two different nodes in a tree that agree on the propositions $p_1,\ldots,p_n$, can be expressed as $\ef(\mydownarrowx\atroot\ef(\neg x\wedge\bigwedge_{i=1}^{n}p_i\leftrightarrow\myatx p_i))$.
\end{remark}
\begin{remark}[Bisimulations]\label{rem:bisimulations}
Bisimulation equivalence is not respected by hybrid branching-time logics: We can distinguish two isomorphic subtrees by naming the root of one of those subtrees. But they respect \emph{hybrid one-bisimulations} defined in \cite{ArecesBM01} and successfully applied in \cite{SchwentickW07} in the linear framework.
\end{remark}

%% file: automata_def.tex
\subsection{Tree Automata}

The following basic notions about B\"uchi automata on infinite trees are based on the definitions in \cite{Vardi95} and \cite{Vardi98}. For a more general introduction to automata on infinite trees, we refer to \cite{Thomas90}.

A \emph{non-deterministic B\"uchi tree automaton} $A$ is a tuple $(Q,\Sigma,q^0,\delta,F)$, where $Q$ is a finite set of states, $\Sigma$ is a finite alphabet, $q^0\in Q$ is the initial state, $F\subseteq Q$ is a set of final states, and $\delta:Q\times\Sigma\rightarrow 2^{Q^k}$ is a transition function. Whenever $A$ is in state $q$ at a node $x$, it non-deterministically chooses a $k$-tuple $(q_1,\ldots,q_k)$ of states from $\delta(q,V(x))$ and moves to node $x\cdot i$ in state $q_i$ for each $i=1,\ldots,k$.

A \emph{run} $r$ of  $A$ on a $\Sigma$-labeled tree $(T,V)$ is a $Q$-labeled tree $(T,V')$, such that the root is labeled by the initial state and the transition rules are respected, i.e., if a node $x$ is labeled $q$ and its children are labeled $q_1,\ldots,q_k$, then $(q_1,\ldots,q_k)\in\delta(q,V(x))$. A run $r$ is \emph{accepting} if $lim(\pi)\cap F\neq\emptyset$ for every infinite path $\pi$ of $r$, where $lim(\pi)$ is the set of states occurring infinitely on $\pi$. A labeled tree $(T,V)$ is \emph{accepted} by $A$ if there is an accepting run of $A$ on $(T,V)$. The \emph{language} of $A$ is the set of trees accepted by $A$ and denoted $L(A)$.

\begin{proposition}[\cite{Rabin70}]\label{prop:nondetautomata}
Non-emptiness of non-deterministic B\"uchi tree automata can be decided in quadratic time.
\end{proposition}

Alternating one-pebble B\"uchi tree automata generalize this concept in three ways. First, they are two-way automata, i.e., they can also move upward in the tree. Additionally, they can drop a pebble at a position in the tree and lift the pebble again if they are at the position where the pebble was placed. In other words, these automata can mark a position to find it again after moving away. Finally, they can universally and existentially branch into several independent sub-computations. 

More formally, an \emph{alternating one-pebble B\"uchi tree automaton} is a 
tuple $A=(Q,\Sigma,q^0,\delta,F)$, such that $Q$ is a finite set of states, $\Sigma$ is a finite alphabet, $q^0\in Q$ is the initial state, $F\subseteq Q$ is the set of accepting states, and 
$$\delta:Q\times\Sigma\rightarrow  (Q\times\{\text{drop,lift}\})\cup\mathcal{B}^+([k]\times Q)$$
is a transition function. 

In this definition, we use $[k]:=\{-1,0,1,\ldots,k\}$ to give the direction of a move of the automaton and $\mathcal{B}^+(X)$ to denote the set of positive Boolean formulas over $X$, i.e., formulas built from $X$ by $\wedge$ and $\vee$ including \true and \false. Note that such an automaton can send several subcomputations into the same direction, but does not need to go into every direction.

A {\em configuration} $(q,x,y)\in Q\times D^*\times(D^*\cup\{\bot\})$ of $A$
consists of a state, the current position in the tree, and the position of
the pebble, where ``$\bot$'' means that the pebble is not placed. 

A {\em run} $r$ of $A$ on an infinite labeled $k$-ary tree $(T,V)$ is a possibly
infinite tree $(T',V')$ whose nodes are labeled by configurations of $A$. This tree must be compatible with the transition function. For example, for every node $v\in T'$ labeled
by a state $(q,x\cdot c,y)$,
\begin{itemize}
\item if $\delta(q,V(x\cdot c))=(q',\text{drop})$ and $y=\bot$, then $v$ has a child labeled with $(q,x\cdot c,x\cdot c)$, otherwise, i.e. if $y\neq\bot$, the transition cannot be applied;
\item if $\delta(q,V(x\cdot c))=(q',\text{lift})$, then $v$ has a child $(q',x\cdot c,\bot)$ if $x\cdot c=y$, otherwise the transition cannot be applied;
\item if $\delta(q,V(x\cdot c))=(1,q')\wedge(-1,q'')$, $v$ has children labeled
by $(q',x\cdot c\cdot 1,y)$ and $(q'',x,y)$; 
\item if $\delta(q,V(x\cdot c))=(0,q')\vee(2,q'')$, then $v$ has a child labeled
by $(q',x\cdot c,y)$ or a child labeled $(q'',x\cdot c\cdot 2,y)$. 
\end{itemize}
A run is \emph{accepting} if every infinite path
contains infinitely many configurations with states from
$F$. Acceptance of $A$ is defined as usual.

%% file: expressiveness.tex
\section{Expressivity}\label{sec:expressiveness}

We examine the expressive power of hybrid branching-time logics. By Remark \ref{rem:bisimulations}, these logics are strictly more expressive than their classical counterparts. In the first part of this section, we give two examples where hybrid machinery is used to cover even more expressive classical branching-time logics.
These results are in contrast to \cite{SchwentickW07}, where it was shown that the hybrid version of \ltl is expressively equivalent to \ltl.
The second part compares branching-time logics with the $\myN$-operator to hybrid branching-time logics.

\subsection{Capturing Classical Branching-Time Logics}

Adding hybrid machinery to \bxf results in a strictly more expressive logic.

\begin{theorem}\label{th:expr:hbxf}
\hbxf is strictly more expressive than \bxf.
\end{theorem}
\begin{proof}
It is known from \cite{LaroussinieS95} that the extension of \bxf with one of the past modalities \myS or \myY is strictly more expressive than \bxf. We show that both \myS and \myY can be expressed in \hbxf.

For the \myY modality, the idea is to fix the current node by naming it $x$ and then to jump to the root and move forward to the node where $\ex x$ holds. This node is the unique predecessor of the node named $x$. If the latter node is already the root, no predecessor exists and the following formula evaluates to false.
$$\myY\varphi\equiv \mydownarrowx\atroot\ef((\ex x)\wedge\varphi)$$
The Since-modality can be replaced in a similar way:
$$\varphi\myS\psi\equiv \mydownarrowx\atroot\ef(\ef x\wedge\psi\wedge(x\vee\ex(\ef x\wedge\ag(\ef x\rightarrow\varphi) ))),$$
respecting that $\psi$ either holds at the current or at some previous node.\qed
\end{proof}

As both formulas given in the previous proof are of linear size in the length of the past-formulas, we obtain the following intensification of Theorem~\ref{th:expr:hbxf}.
\begin{corollary}
There is a linear translation from \bxfys to \hbxf.
\end{corollary}

This shows that every hybrid branching-time logic containing \myX and \myF can refer to the past. In particular, the hybrid version of \ctl captures the extension of \ctl with past modalities, which is known to be strictly more expressive than the pure future logic \cite{LaroussinieS95}. As hybrid branching-time logics do not respect bisimulation-equivalence, this inclusion is strict.
\begin{theorem}
\hbxu is strictly more expressive than \pctl.
\end{theorem}

The bisimulation argument even shows that there cannot be a translation from hybrid branching-time logics into \pctlstar, since the latter can express only bisimulation-invariant properties.

On the other hand, we conjecture that it is not possible to express the \ectl-formula $\efinf p$ in \hbxu. In this case, \hbxu and \ctlstar are incomparable with respect to expressive power.

Finally, hybrid branching-time logics are obviously fragments of Monadic Path Logic (\mpl), the fragment of \mso where set-quantification is restricted to paths. This inclusion can be proved to be strict by observing that hybrid branching-time logics respect hybrid one-bisimulations as defined in \cite{ArecesBM01}.

\subsection{Expressing ``From Now On''}\label{sec:N}

The temporal operator \myN for ``from now on'' was introduced by Laroussinie and Schnoebelen to branching-time logics with past \cite{LaroussinieS95}. The semantics of \myN is given by:
$T,n\models\myN\varphi$ iff $T',\varepsilon\models\varphi$,
where $T'=\{m\in\mathbb{N}^*\mid n\cdot m\in T\}$ is the subtree of $T$ rooted at $n$.
That is, \myN allows to \emph{forget about the past}.

In \cite{LaroussinieS95} the authors provide several results on whether \myN adds expressive power to branching-time logics with past. E.g., it does for \bt{\myX,\myF,\myY,\myP} but does not for \pctl and \pctlstar.
Moreover, they argue that \myN offers a more convenient way to describe some properties in branching-time logics with past (see \cite{LaroussinieS95} for an example), which is partially  attributed to the succinctness of logics with the \myN-operator.\footnote{To the best of our knowledge, this succinctness gap has so far only been proved for the case of linear temporal logic in \cite{LaroussinieMS02}. Succinctness and complexity for branching-time logics with \myN seem to be open problems.}

The following proposition shows that hybrid branching-time logics offer at least the same convenience and are at least as succinct as the logics including~\myN.

\begin{proposition}
For every set of modalities $C\subseteq\{\myX,\myF,\myU,\Finf,\myY,\myP,\myS\}$, there is a linear translation from \bt{\myN,C} to \hb{\myP,C}.
\end{proposition}
\begin{proof}
Given a \bt{\myN,C}-formula $\varphi$, we obtain an equivalent \hb{\myP,C}-formula $\mydownarrowx\varphi'$ by substituting \myN by the downarrow-operator and guarding all past modalities. I.e., $\varphi'$ results from $\varphi$ by applying the following rules once for every past modality and every \myN-operator:
\begin{center}
\begin{tabular}{rclrcl}
$\myN\psi$ & $\rightarrow$ & $\mydownarrowx\psi\qquad\qquad$ & $\myP\psi$ & $\rightarrow$ & $\myP(\myP x\wedge\psi)$\\
$\myY\psi$ & $\rightarrow$ & $\myY(\myP x\wedge\psi)\qquad\qquad$ & $\varphi\myS\psi$ & $\rightarrow$ & $\varphi\myS(\myP x\wedge\psi)$
\end{tabular}
\end{center}

Requiring \myP is only a restriction if \myY is the only past modality in $C$.
\qed
\end{proof}

As we show in the next section, we cannot add \myN on top of hybrid branching-time logics without blowing up the complexity of satisfiability non-elementarily. Intuitively, this is because \myN can play the role of a second state variable, therefore enabling us to talk about three points at the same time: the new root created by \myN, the state named $x$, and the current state.
 
But the reader should be warned not to think of \myN as a kind of state variable, since the ability to name a state and then to talk about its past is crucial to most results in this paper.

%% file: satisfiability.tex
\section{Complexity of the Satisfiability Problem}\label{sec:satisfiability}

The main motivation behind the one-variable approache to hybrid branching-time logics is to tame the complexity of the satisfiability problem. This section shows that this approache was successful by providing a \TWOEXPTIME-completeness result for satisfiability of several hybrid branching-time logics.

The proof of the lower complexity bound for \hbxf is by a reduction from the $2^n$-corridor tiling game.
We first define the \emph{$2^n$-corridor tiling problem}. An instance $I=(T,H,V,n)$ of
this problem consists of a finite set $T$ of tile types, horizontal and
vertical constraints $H,V\subseteq T\times T$, and a number $n$ given
in unary. The task is to decide, whether $T$ tiles the
$2^n\times m$-corridor for some $m$, respecting the constraints $H$ and $V$ and some border constraints, especially on the top row to be
reached.

The \emph{$2^n$-corridor tiling game} is played by two players $E$ and $A$ on an instance $I$ of the $2^n$-corridor tiling problem. The players alternately place tiles starting with player $E$ and following the constraints $H$ and $V$, as the opponent wins otherwise. $E$ wins the game if the required top row is reached. To decide whether $E$ has a winning strategy in such a game is complete for \TWOEXPTIME \cite{Chlebus86}.

\begin{proposition}\label{prop:sathard}
Satisfiability of \hbxf is hard for \TWOEXPTIME.
\end{proposition}
\begin{proof}
Let $I=(T,H,V,n)$ be an instance of the $2^n$-corridor tiling
problem. We build an \hbxf-formula $\varphi_I$ of size polynomial
in $|I|$ that is satisfiable if and only if player $E$ has winning strategy in the tiling game on $I$.

Such a winning strategy is a finite $T$-labeled tree whose levels alternately correspond to moves of $E$ and $A$.
A node corresponding to a move of $E$, as the root for example, has one child for every possible next move of $A$. Nodes representing moves of $A$ have only one child: \emph{the best move $E$ can make}. In order for the strategy to be winning, every path in this tree has to correspond to a correct tiling reaching the required top row.

The formula $\varphi_I$ consists of two parts. The first part
describes an encoding of a winning strategy, using a numbering of the states belonging to one row of the tiling as shown in
Figure \ref{fig:tiling}. Numbers are encoded by $n$ propositions, one
for each digit. The second part basically contains the conditions posed by $H$ and $V$.

\begin{figure}[b]
\footnotesize
\begin{center}
\begin{pspicture}(0,-0.2)(12.0,0.6)

\cnode(0,0){1mm}{a0}

\cnode(1,0){1mm}{a1}
\pnode(1.5,0){a2}
\rput(1.75,0){$\cdots$}
\pnode(2.0,0){a3}
\cnode(2.5,0){1mm}{a4}

\cnode(3.5,0){1mm}{a5}
\pnode(4.0,0){a6}
\rput(4.5,0){$\cdots$}
\pnode(5.0,0){a7}
\cnode(5.5,0){1mm}{a8}

\cnode(6.5,0){1mm}{a9}
\pnode(7.0,0){a10}
\rput(7.25,0){$\cdots$}
\pnode(7.5,0){a11}
\cnode(8,0){1mm}{a12}

\cnode(9,0){1mm}{a13}
\cnode(10,0){1mm}{a14}
\cnode(11,0){1mm}{a15}
\pnode(11.5,0){a16}
\rput(11.8,0){$\cdots$}

\ncline{->}{a0}{a1}
\ncline{->}{a1}{a2}
\ncline{->}{a3}{a4}
\ncline{->}{a4}{a5}
\ncline{->}{a5}{a6}
\ncline{->}{a7}{a8}
\ncline{->}{a8}{a9}
\ncline{->}{a9}{a10}
\ncline{->}{a11}{a12}
\ncline{->}{a12}{a13}
\ncline{->}{a13}{a14}
\ncline{->}{a14}{a15}
\ncline{->}{a15}{a16}
\psframe(0.7,-0.2)(2.8,0.2)
\psframe(6.2,-0.2)(8.3,0.2)

\rput(0,-0.4){$root$}
\rput(1.75,-0.4){row 1}
\rput(7.25,-0.4){row $m$}
\rput(0.0,0.4){$q_{\#}$}
\rput(3.5,0.4){$q_{\#}$}
\rput(5.5,0.4){$q_{\#}$}
\rput(9.0,0.4){$q_{\#}$}
\rput(10,0.4){$q$}
\rput(11,0.4){$q$}
\rput(1.0,0.4){$1$}
\rput(2.5,0.4){$2^n$}
\rput(6.5,0.4){$1$}
\rput(8.0,0.4){$2^n$}

\end{pspicture}
\end{center}
\caption{A path in the encoding of a winning strategy for the $2^n$-corridor tiling game with $m$ rows.}
\label{fig:tiling}
\end{figure}
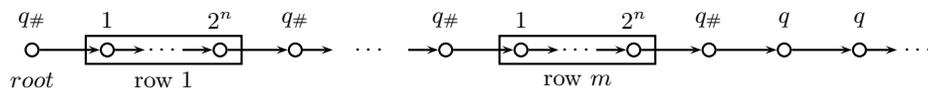

To make this more precise, $\varphi_I$ is the conjunction of the two formulas $\varphi_{struc}$ and $\varphi_{tiles}$. The first formula, $\varphi_{struc}$, starts by separating the lines of the tiling by an additional state labeled by the proposition symbol $q_{\#}$ and marking the states beyond the encoding by the proposition symbol $q$.
\begin{eqnarray*}
	\varphi_{struc} & = & root\wedge q_{\#}\wedge\neg q\wedge(\bigwedge_{i=0}^{n-1}\neg q_i)\wedge \ag((\neg q\wedge \neg q_{\#})\vee((q\leftrightarrow\neg q_{\#})\wedge\bigwedge_{i=0}^{n-1}\neg q_i))\\
	&& \wedge\af(q_{\#}\wedge\ag q)\wedge\ag(q\rightarrow\ag q) \wedge\varphi_{num}
\end{eqnarray*}

To get a correct numbering of the states representing one row of the tiling, $\varphi_{num}$ requires every state to have only properly numbered direct successors. This numbering is required to check the vertical constraints.
\begin{eqnarray*}
	 \varphi_{num} & = & \ag([q_{\#}\rightarrow(\ax(\neg q\wedge\neg q_{\#}\wedge\bigwedge_{i=0}^{n-1}\neg q_i)\vee\ax q)]\\
	 &&\qquad\;\wedge[(\neg q\wedge \neg q_{\#})\rightarrow((\bigwedge_{i=0}^{n-1}q_i\wedge\ax q_{\#}) \vee\nu)])\\
	 \nu & = & \mydownarrowx\ax\bigvee_{i=0}^{n-1}(\bigwedge_{j<i}(q_j\leftrightarrow\myatx\neg q_j)\wedge q_i\wedge \myatx \neg q_i\wedge\bigwedge_{j>i}(q_j\leftrightarrow\myatx q_j))
\end{eqnarray*}

The second part of $\varphi_{I}$ expresses that every state corresponding to a move of one of the players is labeled by exactly one tile, using proposition symbols $p_t$ to represent the tiles, that the conditions in $H$ and $V$ are respected, and that all possible moves of $A$ are represented.
\begin{eqnarray*}
	\varphi_{tiles} & = & \ag([\neg q\wedge \neg q_{\#}]\rightarrow[\bigvee_{t\in T}(p_t\wedge\bigwedge_{t\neq t'\in T}\neg p_{t'})\wedge\theta_H\wedge\theta_V\wedge\theta_A])\\
	\theta_H & = & \neg\bigwedge_{i=0}^{n-1}q_i\rightarrow \bigwedge_{t\in T}(p_t\rightarrow\ax\bigvee_{(t,t')\in H} p_{t'})\\
	\theta_V & = & \mydownarrowx\atroot\ag([\ex\ef x\wedge\neg\ef(q_{\#}\wedge\ex\ef(q_{\#}\wedge\ef x))\\
	& & \wedge\bigwedge_{i=0}^{n-1}(q_i\leftrightarrow\myatx q_i)]\rightarrow\bigwedge_{t\in T}(p_t\rightarrow\bigvee_{(t,t')\in V}\myatx p_{t'})))\\
	\theta_A & = & \neg q_0\rightarrow \bigwedge_{t\in T}(p_t\rightarrow\bigwedge_{(t,t')\in H}[\ex p_{t'}\vee\mydownarrowx\atroot\ag([\ex\ef x\wedge\neg\ef(q_{\#}\\
	& & \wedge\ex\ef(q_{\#}\wedge\ef x))\wedge \bigwedge_{i=0}^{n-1}(q_i\leftrightarrow\myatx q_i)]\rightarrow\ex[\ef x\wedge \bigvee_{(t'',t')\not\in V}p_{t''}])])\\
\end{eqnarray*}

We omit formulas for the border constraints, which are straightforward.\qed
\end{proof}

Before we proceed with the upper bound, we show that the lower bound is due to the succinctness of hybrid formulas.

\begin{theorem}
The succinctness of \hb{\myF} with respect to \ctl is $O(n)!$, i.e., there is a \hb{\myF}-formula of length $O(n)$ such that every equivalent \ctl-formula is at least of length $O(n)!$.
\end{theorem}
\begin{proof}
We consider the \ctlplus-formula $\myE(\myF p_1\wedge\myF p_2\wedge\cdots\wedge\myF p_n)$,
expressing that there exists a path such that each of the propositions $p_1,\ldots,p_n$ holds at some node on the path. Adler and Immerman proved  that one requires a formula of size $O(n)!$ to express this property in \ctl \cite{AdlerI03}.
The following \hbf-formula has only size $O(n)$.
$$\ef(\mydownarrowx\atroot\bigwedge_{i=1}^{n}\ef(p_i\wedge\ef x))$$

The crucial point is that this property depends only on a finite prefix of the path, which can be fixed by naming its last state $x$.
\qed
\end{proof}

The proof of the upper complexity bound for satisfiability of hybrid branching-time logics uses the automata-theoretic approach to branching-time logics (see \cite{Vardi07} and references therein), extended to the hybrid framework.

Before we go on, we observe that nesting of the \mydownarrow-operator can be avoided.

\begin{lemma}\label{lem:nonesting}
For every \hbxufinfys-formula $\varphi$, there is an equivalent
formula $\psi$ of length $O(|\varphi|)$ without nested occurrences
of the $\mydownarrow$-operator. 
\end{lemma}
\begin{proof}
We add, for each sub-formula $\theta=\mydownarrowx\xi$ of $\varphi$, a
new proposition $p_\theta$. In a bottom-up fashion, we replace every
occurrence of a formula $\theta$ by $p_\theta$ and add to $\varphi$
one conjunct $\ag(p_{\theta}\leftrightarrow\theta')$, for every $\theta$.
Here, $\theta'$ results from $\theta$ by replacing all strict sub-formulas
$\mydownarrowx\chi$ by the respective proposition.
\qed
\end{proof}

We can now prove the main theorem of this paper.
 
\begin{theorem}\label{theo:satcomplete}
For every set of temporal operators $C\subseteq\{\myU,\myY,\Finf,\myP,\myS\}$, 
the satisfiability problem for \hb{\myX,\myF,C} is complete for \TWOEXPTIME.
\end{theorem}
\begin{proof}
The lower bound was proved in Proposition \ref{prop:sathard}.

The proof of the upper bound is a extension of a
proof in \cite{Vardi95}, constructing an alternating B\"uchi tree
automaton for a given \ctl-formula. 

Given an \hbxufinfys-formula $\varphi$ without nested occurrences of the 
$\mydownarrow$-oper\-ator, we build an alternating one-pebble B\"uchi tree
automaton $A_{\varphi}=(Q,\Sigma,q^0,\delta,F)$, with
$\Sigma=2^{\prop}$, such that $\varphi$ holds at the root of some $\Sigma$-labeled tree $(T,V)$ if and only if
$A_{\varphi}$ accepts this tree. This reduces the satisfiability problem for \hbxufinfys to non-emptiness of alternating one-pebble B\"uchi tree automata. The latter problem is proved to be in \TWOEXPTIME in Section~\ref{sec:automata}.

In the following, we denote the \emph{dual} of a formula $\psi$ by
$\overline{\psi}$. It is obtained from $\psi$ by switching $\wedge$
and $\vee$, and by negating all other maximal subformulas (we identify $\neg\neg\psi$ with $\psi$), e.g., $\overline{x\vee(\neg
  x\wedge\ef p)}=\neg x\wedge(x\vee\neg\ef p)$
(cf. \cite{Vardi95}). 

The set $Q$ of states is based on the Fisher-Ladner-closure of $\varphi$, consisting of the subformulas of $\varphi$ and their
duals. Additionally, the formula $(\ex\efinf\psi)\wedge\psi$ and all its subformulas are included for every subformula $\efinf\psi$ of $\varphi$. The initial state $q^0$ is $\varphi$. The set $F$ of accepting states contains $\true$ and all formulas of the form $\neg\myE(\chi\myU\psi)$, $\neg\myA(\chi\myU\psi)$, and $(\ex\efinf\psi)\wedge\psi$ from
$Q$. The transition function $\delta$ is defined by induction on the
formula structure:
\begin{center}
	\begin{tabular}{rcl|rcl}
	  $\delta(\true,\sigma)$ & = & $(0,\true)$ & $\delta(p,\sigma)$ & = & $(0,\true)\qquad\text{if }p\in \sigma$\\
	  $\delta(\neg\psi,\sigma)$ & = & $\overline{\delta(\psi,\sigma)}$ & $\delta(\psi\land\xi,\sigma)$ &  = & $(0,\psi)\wedge(0,\xi)$\\
	  $\delta(x,\sigma)$ & = & $(\true,\text{lift})$ & $\delta(\efinf\varphi,\sigma)$ & = & $\bigvee_{i=1}^k(i,\efinf\varphi)\vee(0,(\ex\efinf\varphi)\wedge\varphi)$\\
	  $\delta(\mydownarrowx\psi,\sigma)$ & = & $(\psi,\text{drop})$& $\;\;\delta(\myE(\chi\myU\psi),\sigma)$ & = & $(0,\psi)\vee((0,\chi)\wedge \bigvee_{i=1}^{k}(i,\myE(\chi\myU\psi))$\\
	  $\delta(\ex\psi,\sigma)$ &  = & $\bigvee_{i=1}^{k}(i,\psi)\;$ & $\delta(\myA(\chi\myU\psi),\sigma)$ & = & $(0,\psi)\vee((0,\chi)\wedge \bigwedge_{i=1}^{k}(i,\myA(\chi\myU\psi))$\\
	  $\delta(\myY\psi,\sigma)$ &  = & $(-1,\psi)$ & $\delta(\chi\myS\psi,\sigma)$ & = & $(0,\psi)\vee((0,\chi)\wedge (-1,\chi\myS\psi)$
	\end{tabular}
\end{center}
\noindent where $\sigma\in\Sigma$, $p\in\prop$, and the notion of a dual is extended to 
$\delta$ in the obvious way, e.g., $\overline{\delta(\ex\psi,\sigma)}=\bigwedge_{i=1}^{k}(i,\overline{\psi})$.

The result then follows from Theorem \ref{theo:altmit}. \qed
\end{proof}

This result is optimal with respect to the number of state variables available. We have shown in Section \ref{sec:N} that the \myN-operator can be simulated by a state variable, and therefore be seen as a ``weak'' kind of variable. In the following, we show that adding the \myN-operator to hybrid branching-time logics causes a non-elementary blow-up in complexity.

We have to be a bit careful when adding the \myN-operator to hybrid branching-time logics. First, what is the semantics of a formula of the form $\atroot\psi$ in the scope of an \myN-operator? As the \myN was introduced to forget about the past, the most natural thing is to define that this formula jumps to the new root created my the \myN-operator. While this is minor if past modalities are available, it is the only reasonable choice for pure future hybrid branching-time logics.

The second difficulty is that the state variable might be bound to some state in the past. In order not to unbind the variable, we assume that in this case the assignment is updated to the current state, i.e., to the new root. But this situation does not occur in the following proof.

\begin{theorem}\label{theo:withN}
The satisfiability problem for \hbxfn has non-elementary complexity.
\end{theorem}
\begin{proof}
We give a reduction from the non-emptiness problem for star-free expressions built from union, concatenation, and negation. This problem
is known to have non-elementary complexity \cite{Stockmeyer74}. 
With a string of length $i$ over an alphabet $\Sigma$ we associate a tree whose first $i+1$ nodes have only one child. All states beyond carry the label $q$ as shown in Figure \ref{fig:redRE}. 

\begin{figure}[t]
\footnotesize
\begin{center}
\begin{pspicture}(0,-0.3)(8.0,0.8)
\cnode(0,0){1mm}{a0}
\cnode(1,0){1mm}{a1}
\cnode(2,0){1mm}{a2}
\cnode(3,0){1mm}{a3}
\cnode(4,0){1mm}{a4}
\cnode(5,0){1mm}{a5}
\cnode(6,0){1mm}{a6}
\cnode(7,-0.4){1mm}{a7}
\cnode(7,0.4){1mm}{a8}
\pnode(7.7,-0.6){a9}
\pnode(7.7,-0.2){a10}
\pnode(7.7,0.2){a11}
\pnode(7.7,0.6){a12}
\rput(8,0.4){$\ldots$}
\rput(8,-0.4){$\ldots$}
\ncline{->}{a0}{a1}
\ncline{->}{a1}{a2}
\ncline{->}{a2}{a3}
\ncline{->}{a3}{a4}
\ncline{->}{a4}{a5}
\ncline{->}{a5}{a6}
\ncline{->}{a6}{a7}
\ncline{->}{a6}{a8}
\ncline{->}{a7}{a9}
\ncline{->}{a7}{a10}
\ncline{->}{a8}{a11}
\ncline{->}{a8}{a12}
\rput(0.0,-0.3){$root$}
\rput(1.0,0.3){$p_a$}
\rput(2.0,0.3){$p_b$}
\rput(3.0,0.3){$p_b$}
\rput(4.0,0.3){$p_a$}
\rput(5.0,0.3){$p_a$}
\rput(6.0,0.3){$q$}
\rput(7.0,0.7){$q$}
\rput(7.0,-0.1){$q$}
\end{pspicture}
\end{center}
\caption{The tree used to represent the string $abbaa$ in the proof of Theorem \ref{theo:withN}.\vspace{-4mm}}
\label{fig:redRE}
\end{figure}
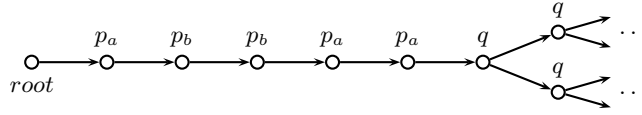

 The following formula $\psi$ holds at the root if and
only if the tree is an encoding of a string, e.g., every state
belonging to the string is labeled by exactly one $p_{\sigma}$. 
\begin{eqnarray*}
  \psi & =  & \ef((q\wedge\ag q)\wedge \mydownarrowx\atroot\ag(\ef\ex x\rightarrow (\neg q \wedge\bigvee_{\sigma\in\Sigma}(p_{\sigma}\wedge\bigwedge_{\sigma\neq\sigma'\in\Sigma}\neg p_{\sigma'}))))\\
  & & \wedge\neg q\wedge \ag(\neg q\rightarrow \ex\mydownarrowx\atroot\ag(\ex x\rightarrow\ax x))
\end{eqnarray*}

\newcommand{\Diamondeq}{\text{\underline{$\myF $}}}

We map every star-free expression $\alpha$ to a formula
\begin{eqnarray*}
	\varphi_{\alpha} & = & \psi\wedge\ef(\neg
        q\wedge \ax q\wedge\mydownarrowx \atroot\alpha'), 
\end{eqnarray*}
where $\alpha'$ is inductively defined as follows:
\begin{center}
\begin{tabular}{p{1cm}rclp{1cm}rcl}
&$\varepsilon'$ & = & $x$ && $\sigma'$ & = & $\ex(x\wedge p_\sigma)\quad\text{, for all }\sigma\in\Sigma$\\
&$\emptyset'$ & = & $ \false$ && $(\alpha\cdot\beta)' $ & = & $\ef(\ef x\wedge\mydownarrowx\atroot\alpha'\wedge\myN \beta')$\\
&$(\neg\alpha)'$ & = & $\neg \alpha'$&& $(\alpha\cup\beta)'$ & = & $ \alpha'\vee \beta'$ \\[2mm]\end{tabular}
\end{center}
The idea is that $x$ is always used to
mark the end of the substring which is matched with respect to a star-free
(sub-)expression while its beginning is at the child of the root ``created'' by \myN.
\qed
\end{proof}

%% file: automata.tex
\section{Non-emptiness of Alternating One-Pebble Tree Automata}\label{sec:automata}

In this section, we show that the non-emptiness problem for
alternating B\"uchi tree automata with one pebble is $\TWOEXPTIME$-complete. 
The proof is based on \cite{SchwentickW07} where $\EXPSPACE$-completeness for the string case is shown.

\begin{theorem}\label{theo:altmit}
Non-emptiness of alternating one-pebble B\"uchi tree automata is
complete for \TWOEXPTIME. 
\end{theorem}

To simplify the presentation of the proof of the upper bound, we assume that we do not have arbitrary positive Boolean combinations on the right-hand side of a transition rule, but only either disjunctions or conjunctions. This is equivalent to the more general notion used before.
A configuration is called existential if the matching transition rule contains a disjunction and universal if it contains a conjunction.

A run $r$ of an alternating  B\"uchi tree automaton $A$ with one pebble
is an infinite in which the nodes are labeled with
configurations $(q,x,y)$, where $q$ is the state, $x$ is a node of the tree and $j$ is the position of the pebble ($\bot$ if the pebble is not placed). We are interested in runs of the following 
structure. A run is {\em homogeneous} if, for every existential configuration $(q,x,y)$, all
nodes of $r$ labeled with configuration $(q,x,y)$ have the same configuration at their child.  

Note that the configuration graph of $A$ on a tree $T$ can be seen as
the arena of a two-player game with a B\"uchi winning condition. Thus, from the
existence of memoryless winning strategies in such games
\cite{EmersonJ91} (see also \cite{Zielonka98}), it follows
that if $A$ has an accepting run on $T$, then it also has a
homogeneous accepting run. 

We show that for each alternating one-pebble B\"uchi tree automaton there is an equivalent
non-deterministic B\"uchi tree automaton of double exponential size. The basic idea is to simulate an accepting homogeneous run of the alternating automaton by running the non-deterministic B\"uchi automaton constructed in \cite{SchwentickW07} for the string case along every branch of the tree.

When arguing about runs, we will make use of the following version of K\"onig's Lemma.
\begin{theorem}[K\"onig's Lemma]
  If in a tree each node has only finitely many children and there
  are nodes of arbitrary depth, then the tree has an infinite path.
\end{theorem}

\begin{proof}[of Theorem \ref{theo:altmit}] Hardness follows from Proposition \ref{prop:sathard} via the translation presented in the proof of Theorem \ref{theo:satcomplete}.

For the upper bound, we show that, for each
alternating one-pebble B\"uchi tree automaton $A$, there is an equivalent
non-deterministic B\"uchi tree automaton $B$ of  size $|\Sigma_A|\cdot2^{2^{O(|Q_A|)}}$,
which can be constructed from $A$ in space exponential in $|Q_A|$. 
The result then follows by Proposition \ref{prop:nondetautomata}. 
As already indicated above, $B$ checks, on input $T$, whether $A$ has a homogeneous
accepting run $r$ on $T$. 

A run is {\em not} accepting if and only if it has a non-accepting
path. Thus, $B$ checks that $r$ has no non-accepting path. 
The non-accepting paths can be classified as follows. First, a path $\pi$
can be {\em bounded}, i.e.,  there is
some $m\in\mathbb{N}$ such that all nodes of $T$ occurring in the labels along $\pi$ are at most
of depth $m$, otherwise, we call $\pi$ {\em unbounded}.   

There are two kinds of unbounded non-accepting paths:
\begin{enumerate}
  \item[(1a)] At some point, the automaton $A$ drops the pebble at some
    node $y$ and never lift it again. In this case, all further
    configurations are of the form $(q,x,y)$, for some $q,x$.
  \item[(1b)] Otherwise, the path has infinitely many configurations of the
    form $(q,x,\bot)$.
\end{enumerate}
In both cases $x$ can be arbitrarily deep in $T$.

Likewise, there are two kinds of bounded non-accepting paths.
\begin{enumerate}
  \item[(2a)] The first kind drops the pebble at some node $y$ and never lifts
    it again. Thus, there are $q$ and $x$ such that $(q,x,y)$ occurs
    infinitely often on $\pi$ and  there is a subpath from 
  configuration $(q,x,y)$ to $(q,x,y)$ which does not visit any accepting
  state, does not lift the pebble, and does not visit any strict descendant of $x$.
\item[(2b)] The other kind of bounded paths has infinitely many
  configurations of the form $(q,x,\bot)$, hence there is again a maximum
  $x$ and a state $q$ such that $(q,x,\bot)$ occurs infinitely often and
  only finitely many nodes have a configurations $(p,x',\bot)$ with
  $x'>x$. Therefore, there is a subpath from
  configuration $(q,x,\bot)$ to $(q,x,\bot)$ which does not visit any accepting
  state and does not visit any descendant of $x$ without having the pebble placed before.
\end{enumerate}
Note that the node $x$ is not unique in both cases. Therefore, $B$ will check the existence of such path at every possible node $x$ of $T$.

In the following we describe the information that $B$ maintains in
order to check that $r$ has only accepting paths.

For each $x$, let $S_x$ be the set of states $q$ for which
$(q,x,\bot)$ occurs in $r$. We consider two kinds of {\em upward paths} ({\em downward paths} are defined accordingly):
\begin{itemize}
\item paths starting from a configuration $(p,x,\bot)$ and ending in a configuration
$(q,x,\bot)$ without an intermediate configuration $(p',x',\bot)$ with
$x < x'$ (intermediate configurations $(p',x',y)$ with
$x\not<y $ and $x<x'$ are allowed);
\item paths starting from a configuration $(p,x,y)$ and ending in a configuration
$(q,x,y)$, without any intermediate lifting of the pebble and without any intermediate configurations $(p',x',y)$ with
$x < x'$.
\end{itemize}

For each node $x$, we denote by $U_x$ the subset of $Q\times Q\times
\{+,\exists\}$, such that
\begin{itemize}
\item $(p,q,+)\in U_x$ if and only if all upward
paths of $r$ from $(p,x,\bot)$ to $(q,x,\bot)$  visit an
accepting state, and 
\item $(p,q,\exists)\in U_x$ if and only if $r$ has a upward
path from $(p,x,\bot)$ to $(q,x,\bot)$,
\end{itemize}
and by $D_x$ the subset of $Q\times Q\times \{+,\exists\}$, such that
\begin{itemize}
\item $(p,q,+)\in D_x$ if and only if all downward
paths\footnote{Since $r$ is homogeneous, all these paths are isomorphic.} of $r$ from $(p,x,\varepsilon)$ to $(q,x,\varepsilon)$  visit an accepting state, and 
\item $(p,q,\exists)\in D_x$ if and only if $r$ has a downward
path from $(p,x,\varepsilon)$ to $(q,x,\varepsilon)$.
\end{itemize} 

It should be noted that in the definition of $D_x$, the actual
position of the pebble does not matter, as long as it is not a descendant of
$x$. The reader should also observe the asymmetry between the
$U_x$ and the $D_x$. The $U_x$ only concern sub-computations from a
configuration without pebble, the $D_x$ only from a configuration with pebble.

Furthermore, $B$ uses the following sets which are
parametrized by the current position $y$ of the pebble.
Let, for each node $x$ and each $y\not > x$,  $S_{x,y}$ be the set of states $p$
such that $(p,x,y)$ occurs in $r$. Likewise, let $U_{x,y}$ be the
subset of $Q\times Q\times \{+,\exists\}$ such that
\begin{itemize}
\item $(p,q,+)\in U_{x,y}$ if and only if all upward
paths of $r$ from $(p,x,y)$ to $(q,x,y)$  visit an
accepting state and 
\item $(p,q,\exists)\in U_{x,y}$ if and only if $r$ has a upward
path from $(p,x,y)$ to $(q,x,y)$.
\end{itemize} 
Recall that upward paths from a configuration $(p,x,y)$ never lift the
pebble. 

Additionally, $B$ uses sets $R_x$, $R_{x,y}$ and $D'_i$ which will be
defined below. For each $x$, we let $X_x$ be the set
$\{(S_{x,y},U_{x,y},R_{x,y}) \mid y\not> x\}$. Finally, for each $x$,
let the characteristic vector $C_x$ of position $x$ be
$(S_x,U_x,D_x,D'_x,R_x,X_x)$. The intended state of $B$ at node
$x$ is basically $(C_{x\cdot -1},C_x)$. 

The sets of the form $S_x,S_{x,y},D_x$ and the transitions of $A$ are
guessed by $B$ and the remaining information can be determined from it.
It is not hard to check that local consistency of these sets can be
 tested by $B$. It should be noted that the computation of
$U_x$ uses $D_x$ to handle subpaths that drop the pebble outside of the subtree rooted at $x$ and
lift it sometime later.

Whether a path of type (2b) exists from node $x$ can be 
inferred from $U_x$ and the transitions $\delta_{q,x}$. Likewise, paths of type (2a) can be tested with the help of the
sets $U_{x,y}$. 

Thus, it remains to describe how to rule out paths of types (1a) and
(1b) and how to check that the sets $D_x$ are correct. 

To this end, we define for every path $\pi$ in $T$ an increasing sequence $l_0,l_1,\ldots$ of
nodes of $\pi$ as follows. First of all,  $l_0=\varepsilon$. Given $l_k$,
 $l_{k+1}$ is the minimal node $l>l_k$ on $\pi$ such that the
following conditions hold.
\begin{enumerate}[(i)]
\item For each state $q\in S_{l_k}$,
  each subpath of $r$ starting from a node with configuration
  $(q,l_k,\bot)$ and reaching a configuration $(p,l,\bot)$ 
  contains an accepting state.
 \item For each $y\not >l_k$ and each state $q\in S_{l_k,y}$,
  each subpath of $r$ starting from a node with configuration
  $(q,l_k,y)$, reaching a configuration $(p,l,y)$ without lifting the pebble
  contains an accepting state.
\item For each $(p,q,\exists)\in D_{l_k}$, there is a path in $r$ from
  $(p,l_k,\varepsilon)$ to $(q,l_k,\varepsilon)$ on which no node is a descendant of $l$.
\item For each $(p,q,+)\in D_{l_k}$, all paths in $r$ from
  $(p,l_k,\varepsilon)$ to $(q,l_k,\varepsilon)$ contain an accepting state and do not visit any descendant of $l$.
\end{enumerate}

With the help of K\"onig's Lemma, it is not hard to see that such an $l$ exists if
$r$ is accepting. 

For each $k$ and each $x$ with $l_k<x\le l_{k+1}$, let $R_x$ be the set of
states $q$ such that there is a node of $r$ labeled with configuration $(q,x,\bot)$ that can be reached from a configuration $(p,l_k,\bot)$ for some $p\in S_{l_k}$, without passing any accepting state and without visiting any descendant of $x$. Note that $R_{l_{k+1}}=\emptyset$ by the definition of $l_{k+1}$. 

Likewise, for each $x,y$, $y\not >
x$, let  $R_{x,y}$ be the set of
states $q$ such that there is a node of $r$ with configuration $(q,i,j)$
 that can be reached from a configuration $(q',l_k,j)$ to $w$ , without passing any accepting state,without lifting the pebble, and without visiting descendants of $x$. Again by the definition of
$l_{k+1}$, $U_{l_{k+1}}=\emptyset$. 

Finally, let $D'_x$ be a
set of tuples $(p,q,\exists)$ and $(p,q,+)$ from $D_i$, which still have to be
fulfilled in order to satisfy conditions (iii) and (iv) for $k$.

The accepting states of $B$ are those for which $R_x=\emptyset$, for
all $(S,U,R)\in X_x$, $R=\emptyset$, and $D'_x=\emptyset$. 

It is not hard to see that $B$ can maintain the characteristic vectors
$C_x$ and that $B$ accepts $T$ if and only if $A$ has a homogeneous
accepting run on $T$.

Furthermore, there are at most
doubly exponentially many different possible sets $X_x$ and thus the
number of possible states of $B$ is at most doubly 
exponential in the size of $Q_A$. Using standard space saving techniques,  $B$
can be constructed in space $|\Sigma_A|\cdot 2^{O(|Q_A|)}$.\qed
\end{proof}

%% file: conclusion.tex
\section{Conclusion}\label{sec:conclusion}

We have shown how to extend branching-time logics with hybrid machinery without blowing up complexity non-elementarily. The key to this result was the restriction to a single state variable proposed in \cite{SchwentickW07}.

We studied the satisfiability problem for the hybrid versions of several branch\-ing-time logics, ranging from \ub to \ectl+Past. We proved \TWOEXPTIME-completeness of the satisfiability problem in all cases. The lower bound was additionally explained by the succinctness of hybrid branching-time logics.

To abtain the upper complexity bound, we extended the automata-theoretic approache to hybrid branching-time logics: We proved non-emptiness of  alternating one-pebble B\"uchi tree automata to be \TWOEXPTIME-complete.

We want to give some open problems and directions for further research.
\begin{itemize}
	\item There are a lot of open problems concerning the expressive power of hybrid branching-time logics. E.g., is \hbxf a strict fragment of \hbxu?
	\item We only considered satisfiability, leaving out the model-checking problem. This gap has to be filled in future work.
	\item We restricted to \ctl-like branching-time logics, not allowing Boolean combinations and nesting of temporal operators inside a path-quantifier. Extending our results to such logics is a challenging problem. In particular, the complexity and expressiveness of hybrid \ctlstar should be investigated.
	\item On the purely automata-theoretic side, the result on one-pebble tree automata should be extended to $k$-pebble tree automata. 
\end{itemize}